# Easy/Hard-Transition in k-SAT

Bernd R. Schuh

Dr. Bernd Schuh, D-50968 Köln, Germany; bernd.schuh@netcologne.de

***Abstract.***
A heuristic model procedure for determining satisfiability of CNF-formulae is set up and described by nonlinear recursion relations for *m* (number of clauses), *n* (number of variables) and clause filling *k*. The system mimicked by the recursion undergoes a sharp transition from bounded running times (="easy") to uncontrolled runaway behaviour (="hard"). Thus the parameter space turns out to be separated into regions with qualitatively different efficiency of the model procedure. The transition results from a competition of exponential blow up by branching versus growing number of orthogonal clauses.

***Introduction***.

Complexity classes in general differ by their optimal running times on a classical Turing computer. For all problems in P, e.g., algorithms with polynomial running time (p.t.) can be found, whereas – unless P=NP – at least some problems in NP have exponential running times under any procedure whatsoever. (Note, that this is not a definition of complexity classes P and NP, but one criterion to distinguish them.) Problems with p.t. running times are classified as "easy". For practical purposes it is important to identify so called "hard" problems which prove difficult to be solved by a variety of algorithms. They might also be theoretically fruitful in leading the way to NP-problems which cannot be solved by any p.t.-algorithm, thus proving P≠NP. One criterion in use for locating "hard" versus "easy" SAT problems is the phase transition appearing in random k-SAT. There seems to be a sharp transition between satisfiable and non-satisfiable propositional formulae along the line $m/n$ = const. ≈ 4.26 for *k*=3 [1]. Other statistical models predict a similar phase transition for unrestricted random

SAT (i.e. $k$ not fixed), in the neighbourhood of the line $m/n \approx ln(2)(1-\kappa/2n)^{-n}$, where $\kappa$ is the mean number of literals per clause[2]. Hard formulae are expected mainly in the neighbourhood of such lines. It is not absolutely clear, why, because "hardness" is a property defined via the effectiveness of solving procedures. Why should those be sensitive to the satisfiability of formulae? Indeed I will show that other regions in problem space can be identified as critical in the sense that the effectiveness of the solving procedure undergoes a transition from "easy" to "hard". Here "easy" does not necessarily mean p.t. and "hard" not necessarily exponential. The point is that efficiency measured by any reasonable running time undergoes a dramatic jump when crossing borders between regions. It is not absolutely clear whether this is a property of the solving procedure, the problem class or both – a doubt one can raise in many such arguments, however.

### *Algorithm.*

My starting point is a SAT algorithm, which is a classical branching procedure, that takes into account both truth values for each variable. It has been described as "splitting" elsewhere, [3], and probably is known under different names for SAT solver specialists. I will call it SPLIT in the following. The procedure aims at deciding the question whether a formula is satisfiable or not. It does not aim at finding all or one solution explicitly.

The algorithm is easiest described as follows: let F denote a CNF-formula with $m$ clauses and $n$ variables. We write $F(a_1, a_2, ..., a_n|m)$ to indicate the dependence on the $n$ variables and the number of clauses. The procedure is as follows:

Step 1: Define $G := F(a_1, a_2, ..., a_n|m) \vee F(\bar{a}_1, a_2, ..., a_n|m)$.

Step 2: Rewrite G as a CNF-formula $F'(a_2, a_3, ..., a_n|m')$.

Step 3: Repeat the procedure for each remaining variable.

Obviously, SPLIT consists of a finite number of steps and leads to a conclusive result. After n steps it ends in TRUE, a tautology, or FALSE, empty clauses, corresponding to satisfiability or non-satisfiability. Each step consists of at most p.t. steps in the length of the respective formula. The procedure is not necessarily a p.t. process, however, because the number of clauses may increase considerably in each step, and in sum exponentially or even overexponentially in $m$ or $n$. To be more precise, imagine F decomposed in a part $F_1$ which contains literals $a_1$ and $\bar{a}_1$, and a part $F_R$ which does not. Then

$$G = F(a_1, a_2, ..., a_n|m) \vee F(\bar{a}_1, a_2, ..., a_n|m)$$

$$= \{F_1(a_1, a_2, ..., a_n|m_1) \vee F_1(\bar{a}_1, a_2, ..., a_n|m_1)\} \wedge F_R(a_2, ..., a_n|m_R)$$

Let $F_1$ consist of $r$ clauses $a_1 \vee Y_i$ containing all $a_1$ and $s$ clauses $\bar{a}_1 \vee Z_j$ containing all $\bar{a}_1$, then G simplifies to

(1) $\qquad G = \{(Y_1 \wedge ... \wedge Y_r) \vee (Z_1 \wedge ... \wedge Z_s)\} \wedge F_R(a_2, ..., a_n|m_R)$

Variable $a_1$ is completely eliminated now. To repeat the procedure with the next variable the term in {}-brackets must be transformed into CNF form ("CNF-isation"):

(2)  $(Y_1 \wedge ... \wedge Y_r) \vee (Z_1 \wedge ... \wedge Z_s) = Y_1 \vee Z_1 \wedge Y_1 \vee Z_2 ... \wedge Y_1 \vee Z_s \wedge Y_2 \vee Z_1 \wedge ... Y_i \vee Z_j ... \wedge Y_r \vee Z_s$

Thus the new F' consists of $m' = rs+m_R$ clauses, in principle. Not only the product *rs* causes trouble, but also the fact that *r* and *s* may increase in each step of the process, because CNF-isation generates more clauses with the remaining $a_i$ than there were before. Consider the worst case $r=s=m/2$. Then the number of clauses grows overexponentially fast, because after j steps the number of clauses has blown up to $4(m/4)^{2^j}$ already. There are mechanisms, however, to reduce the number of emerging clauses efficiently, i.e. in p.t. intermediate steps. The most important is

> (i) orthogonal clauses :    If $Y_i \vee Z_j$ is a tautology (i.e. it contains a variable b and its complement $\overline{b}$ ) it need not be counted.

Additionally in the course of CNF-isation several identical clauses will turn up, which need to be counted once only. Removing identical clauses is a special case of removing

> (ii) redundant clauses:    If, in set theoretical notation, a clause is contained in another one, A $\subseteq$ B, then B can be neglected in the conjunction, because it does not influence satisfiability.

Whereas mechanism (i) leads to a clear reduction of clauses in each step, we will see that (ii) may have an adverse effect, because it can diminish the appearance of orthogonal clauses. Other p.t. simplifications which may be performed in each step are mentioned in [3].

Splitting may not be the most clever choice in practice because it throws away the information which truth value is the correct one for each variable in each step, but it is a straight forward decision algorithm to discern satisfiability from non-satisfiability. Thus its running time measures its efficiency with respect to solvability only. A reasonable measure for the performance of the procedure may be any power of *m* or *mn* or any combination thereof. Thus we need to answer the question: How does *m* vary quantitatively in the iteration process, if one or both attenuation effects are taken into account? To answer that question I will set up a heuristic model for the variation in clause number and investigate its consequences.

### *Equations.*

Of the two attenuation mechanisms only the first, orthogonality will be taken into account in the following. Let F be an arbitrary CNF-formula with *m* clauses and *n* variables. In general, each clause contains $k_i$ literals (variable $a_i$ or its complement $\overline{a}_i$ ). Each variable (and its complement) appears $p_i$ times in the conjunction of clauses. I will call $p_i$ the appearance of variable $a_i$. Obviously

$$\sum_{\text{clauses}} k_j = \sum_{\text{variables}} p_i ,$$ a conservation theorem for the total number of literals in the formula.

Even if one starts with a homogeneous F, where all $k_i=k$ and $p_i=p$, SPLIT in general leads to intermediate formulae with inhomogeneous distribution of literals. To make things tractable I model the splitting process by using idealized, homogeneous formulae in each step which are described by the mean number of literals $k = \sum_{\text{clauses}} k_j /m$ and the mean appearance $p = \sum_{\text{variables}} p_i /n$. From the conservation theorem we have

(3)     $x := p/m = k/n$.

Thus three parameters, e.g. *m, n, k*, suffice to characterize the process. Furthermore I assume symmetry, i.e. for each variable the number of positive and negative literals is the same, namely *p*/2. Thus our starting point is a formula F with *m* clauses, n variables and $p/2 = mk/2n$ many $a_1$ and *p*/2 many $\bar{a}_1$ (*r=s* in the notation of equ. (2)). Now we can calculate the number of clauses *m'* from (2). Without taking into account any reduction mechanism we would have

$$m' = (p/2)^2 + m_R = p^2/4 + m-p = m(mx^2/4 + 1-x)$$

for the number of clauses of the emerging F'.

However, the term $p^2/4 = (p/2)(p/2)$ is a gross overestimate, because whenever in the process of CNF-isation two orthogonal clauses meet they unite to TRUE and need not be counted. To quantify this observation we focus on the next variable to be eliminated, say $a_2$. How many clauses with $a_2$ and $\bar{a}_2$ do the *r = p*/2 clauses of the Y-block in equ. (1) contain? Now I make use of the assumption that F from the start (and in any intermediate step) is symmetric and homogeneous. Since *x=p/m* is the overall appearance per clause we will have *xp*/2 many literals $a_2$ and $\bar{a}_2$ in this block, and *p*/2-*xp*/2 = (1-*x*)*p*/2 many clauses without this variable. Symmetry implies that we find *xp*/4 many $a_2$-clauses and *xp*/4 many $\bar{a}_2$-clauses in Y-block. The same is true for Z-block, per symmetry. The situation is illustrated in fig. 1.

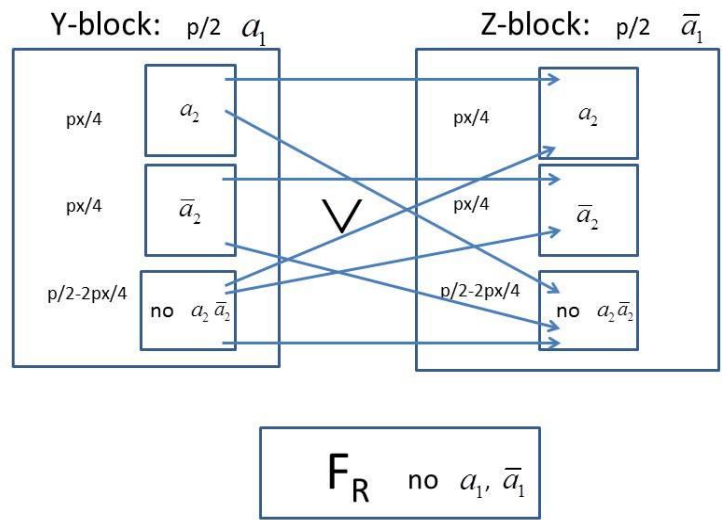

*Figure 1*   Schematic representation of blocks of clauses ready for CNF-isation. Arrows indicate clausewise disjunction.

5Now one can count the nontrivial emerging clauses. Instead of $rs = p^2/4$ one gets:

$$2(\frac{xp}{4})^2 + 4\frac{xp}{4}(1-x)\frac{p}{2} + \frac{p^2}{4}(1-x)^2 = \frac{p^2}{4}(1-\frac{x^2}{2})$$

Thus by first order orthogonality the number of emerging clauses is attenuated by a factor $(1-x^2/2)$ already. In addition, the $a_2$- and $\bar{a}_2$- subblocks contain other variables themselves, which reduce their contribution, e.g. $(xp/4)^2$, by a further factor $1-x^2/2$, and so on. This higher order attenuation is limited by the mean number of literals per clause, $k$. Since in our consideration two variables are used already, namely $a_1$ and $a_2$, the total additional attenuation is given by $(1-x^2/2)^{k-2}$. The final result for $m$ reads

(4) $\quad m' = m - p + \frac{p^2}{4}(1-\frac{x^2}{2})r(n,x) \quad$ with $\quad r(n,x) = \left(1-x^2/2\right)^{nx-2}$

How does $p$ change in the process? The next variable $a_2$ will have an appearance $p'$ which consists of the $x(m-p)$ many $a_2$ and $\bar{a}_2$ in $F_R$ plus the number arising from the CNF-isation of Y- and Z-blocks. Since only the $(1-x)^2 p^2/4$ many clauses emerging from the CNF-isation do <u>not</u> contain any $a_2$ or $\bar{a}_2$, we get instead of the equation preceding equ. (4):

$$2(\frac{xp}{4})^2 + 4\frac{xp}{4}(1-x)\frac{p}{2} = \frac{p^2}{4}(2x - \frac{3}{2}x^2)$$

for the number of clauses with $a_2$ and $\bar{a}_2$, generated by CNF-isation. Also for these terms one has the attenuation by additional variables via a factor $r(n,x)$. Finally:

(5) $\quad p' = x(m-p) + \frac{p^2}{4}(2x - \frac{3}{2}x^2) r(n,x)$

Also $n$ changes, but simply by 1 in each step

(6) $\quad n' = n - 1$

The filling $k$ can be related to the change in $x$ and $n$ via equ. (3), the conservation of literals:

(7) $\quad k' = x'n' = n'p'/m'.$

Equations (3) - (7) are recursion relations stating how the parameters $(m,n,k)$ of a CNF-formula F change under SPLIT. Since the resulting F' is of the same form as F, with different parameters $(m',n',k')$, the process can be repeated until this (simplified) splitting procedure stops at $n'=2$ (modeling requires $k \geq 2$). What can be said about the running time of this model process?

*Iteration.*

One may rewrite the recursion relations as relations for *m*, *n* and the mean filling factor of a clause $x=p/m=k/n$.

(8a) $\quad m' = m(1-x+\dfrac{mx^2}{4}r(n,x)(1-\dfrac{x^2}{2}))$

(8b) $\quad x' = x\dfrac{1-x+\dfrac{mx^2}{4}r(n,x)(2-\dfrac{3}{2}x)}{1-x+\dfrac{mx^2}{4}r(n,x)(1-\dfrac{x^2}{2})}$

(8c) $\quad n' = n - 1$

These nonlinear relations are to be iterated, starting with a given $n_0$, $m_0$ and $k_0$. Since $p \leq m$ we have $x \leq 1$ in each step. Also $x'/x \geq 1$ can be read off directly. Thus *x* grows during the process and can reach its maximum $x^*=1$, which is the only fixpoint for $0 < x \leq 1$. From (8a) it becomes clear that the behaviour of *m* depends on whether the factor in brackets, $m'/m$, is smaller or larger than 1.

The actual iteration is limited by $n_0$, the starting value for the number of variables; iteration stops after $j = n_0 - 2$ steps. When *x* approaches the fixpoint *m* may blow up, exceed $2^{n_0-j}$ (which is a solution for x=1) and grow indefinitely, corresponding to a fixed point $m=\infty$. This runaway behaviour happens if $m_0/n_0$ is sufficiently large. Then also running times will blow up uncontrollably. This behaviour I interpret as indicating that problems in that region of parameters are "hard" to solve. There are starting configurations (i.e. CNF-formulae), however, for which a completely different behaviour becomes possible: For sufficiently small $m_0/n_0$ *m* (and *p*) first grow, then reach a maximum to decrease afterwards, tending quickly to zero, with *x* as quickly saturating at some pseudo fixed point $\bar{x} < 1$. $m$ then decreases according to $\sim \exp((n_0 - j)\bar{x} \ln(1 - \bar{x}^2/2))$ in the course of the iteration. The situation is illustrated in fig. 2 for $k_0=4$, $m_0= 100$ and $n_0= 60$ as starting values. For such values our model test of satisfiability can be stopped when $m_j < 1$. Then a conclusion on solvability is possible long before the final step $j = n_0-2$ is reached. We will consider this behaviour as "easy" in the following, irrespective of the actual values of running times. When $m_0/n_0$ exceeds a critical value, however, the behaviour is dominated by the branching character of the procedure which tends to square $m$ in each step. As a result the iteration produces overflow. To give an impression of the sudden change in character, fig. 3 shows the jump in running time, defined by the sum of $m_j^2 n_j$, for $k=3$ and $m=388$ when $n$ crosses the critical value $n_c=100$. In the right part of the picture the points represent the total running time, i.e. the summation extends to the iteration point where $m_j < 1$, whereas the points in the left part correspond to the summation up to the iteration step when $m_j$ exceeds a given bound ($m_0 2^{n_0}$ e.g.) for the first time to grow indefinitely afterwards.

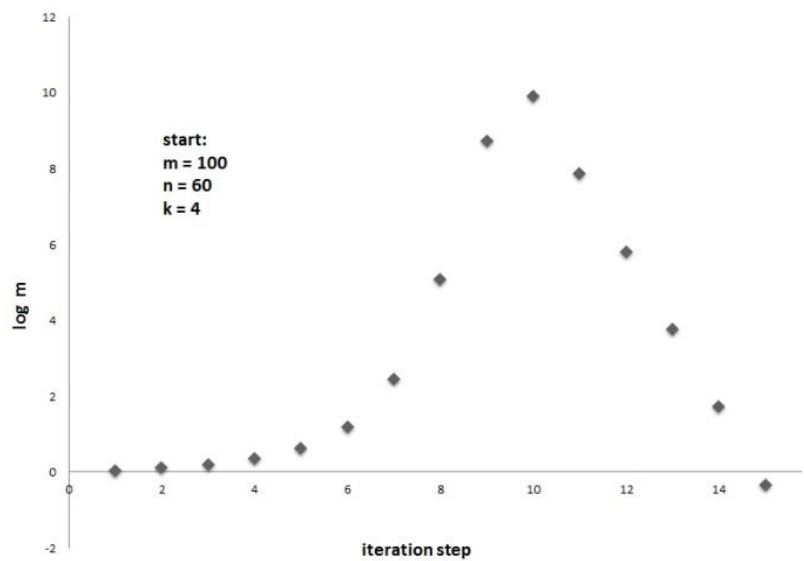

*Figure 2: Iterated number of clauses $m_j$ in the course of the iteration.*

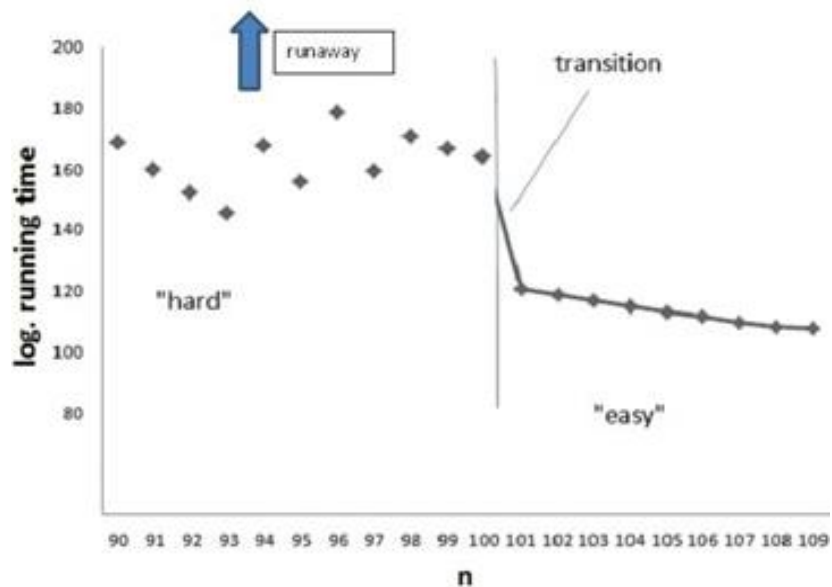

*Figure 3*: Running time as explained in the text near the critical point $n_c=100$, $m_c=388$ for $k=3$.

Fig. 4 shows the transition curve in the *m-n*-plane which separates regions with runaway behavior and "easy" behaviour for $k=3$. Apparently the transition curve is not linear, a power law with exponent ≈ 1.95 produces a good fit. "Easy" problems lie below the curve, "hard" ones above. Note that I use the term "easy" in contrast to the runaway behaviour beyond the transition line, not as an indication of p.t. behaviour. In fact the number of clauses $m_j$ just below the transition line, summed up during the iteration scales like $2^{n_0}$ indicating a "typical" $m_j$ of the order of $\binom{n_0}{j}$.

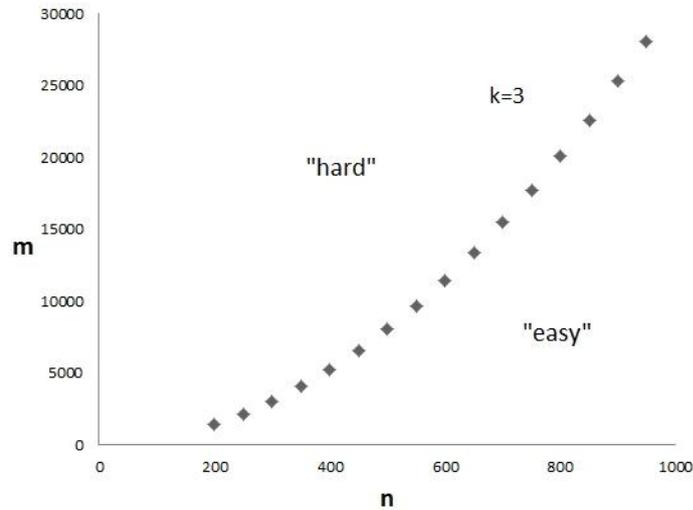

*Figure 4*: *Critical line separating hard and easy problems for k=3.*

The location of the critical line depends strongly on the filling factor *k*. The larger *k* the steeper the critical curves and the larger the "easy" region. This is reasonable, since for large *k=nx* the damping factor *r(n,x)* surpresses the nonlinear quadratic terms in the equations from the beginning and limits the chance for *p* and *m* to blow up sufficiently fast. Or stated in terms of clauses: the larger *k* from the start the greater the chance that clauses are orthogonal and eliminate each other in the CNF-isation early enough to prevent blow up. The dependence of the critical *m* on filling *k* is shown in fig. 5 for *n*=300.

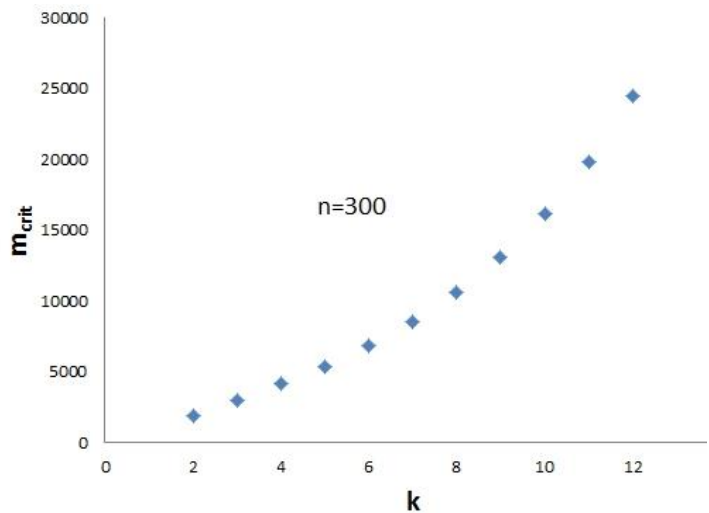

*Figure 5*: *Variation of the critical point $m_c$ with filling factor k for $n_c$=300.*

### *Influence of redundant clauses.*

Attenuation by redundancy (see mechanism (ii) in the introduction), has not been taken into account so far. Let us consider the special case of elimination of identical clauses first. In the framework of this model it could be incorporated via sophisticated statistical arguments. Its effect, however, is clear: it would lead to a further damping of the runaway behaviour which is a characteristic of hardness and it would thus increase the region of "easy" behaviour in problem space. I checked the incorporation of this mechanism schematically by introducing a phenomenological factor $\alpha < 1$ in front of the CNF-isation terms in the recursion relations for *m* and *p*. The only effect is a forseeable shift of the transition curves into the expected direction.

Removing redundant clauses in an intermediate step changes the game once more. The interesting possibility arises that *p* and *m* are attenuated differently. The reason is that redundant clauses contain more literals than the remaining clauses and thus contribute more to the mean *p*. Since they are removed the remaining *p* is attenuated stronger than *m*. To be more precise suppose the CNF-isation step has been performed. It results in $m_{new} = \frac{p^2}{4}(1-\frac{x^2}{2})^{k-1}$ new clauses with appearance $p_{new} = \frac{p^2}{4}(2x-\frac{3}{2}x^2)(1-\frac{x^2}{2})^{k-2}$ (see equations (4) and (5) ). The mean filling of these clauses will be $k_{new} = (n-1)p_{new}/m_{new}$. Next assume that a fraction $\alpha$ of the new clauses can be removed because they are redundant. We are interested in the value of *p* and *m* after the redundant clauses have been removed, $p_{rem}$ and $m_{rem}$. Only those values count for step 3 of the algorithm. *m* is reduced by a factor 1 - $\alpha$ : $m_{rem} = (1-\alpha)m_{new}$. What about *p*? Observe that the redundant clauses will have a larger filling $k_{red}$ than the average $k_{new}$ (that is the essence of redundancy). Let us introduce a second phenomenological constant $\lambda \geq 1$ via $k_{red} = \lambda k_{new}$ and with $\alpha k_{red} + (1-\alpha)k_{rem} = k_{new}$ or likewise $p_{red} + p_{rem} = p_{new}$ it follows that $k_{rem} = (1-\alpha\lambda)k_{new}/(1-\alpha)$ , thus $1 \leq \lambda \leq 1/\alpha$ (because $k_{rem}$ is smaller than $k_{new}$ and positive) and for the appearance of the remaining clauses we get $p_{rem} = (1-\alpha\lambda)p_{new}$ .

Consequently, instead of equations (4) and (5) we end up with

$$m' = m - p + (1-\alpha)\frac{p^2}{4}(1-\frac{x^2}{2})r(n,x) \quad \text{and}$$

$$p' = x(m-p) + (1-\alpha\lambda)\frac{p^2}{4}(2x-\frac{3}{2}x^2)r(n,x)$$

The main point of these considerations was to show, that indeed redundancy creates different weights for the attenuation of *m* and *p*. Even this simplistic model leads to an interesting complication in the recursion relations, because now also *x'*/*x* < 1 becomes possible. As a result the effects of $\alpha$ and $\lambda$ are adverse, because with decreasing *x* the attenuation factor *r(n,x)* increases. Removing identical clauses only ($\alpha > 0$, $\lambda = 1$) leads to a strong damping of runaway effects, as mentioned before. As soon as $\lambda$ is switched on, however, the damping is reduced, leading to a shift of transition curves to lower values of $m_{crit}$, thus diminishing the "easy" area of problem space. As an illustration see fig. 6.

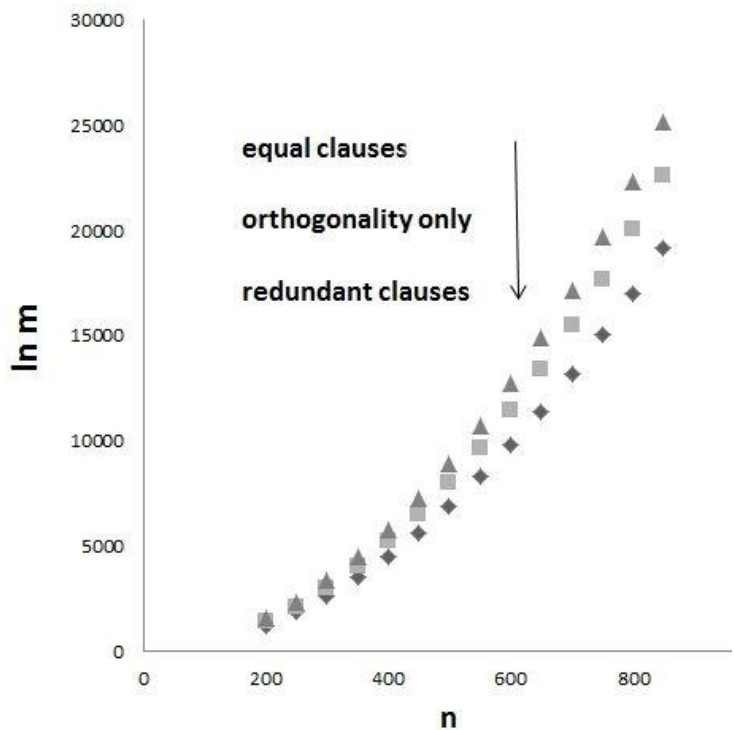

*Figure 6*: *Efficiency transition lines for k = 3 and different attenuation effects. Middle curve: $\alpha = 0$, orthogonality only (as figure 3). Upper curve: Additionally removed identical clauses, $\alpha = .1$, $\lambda = 1$. Lower curve: Redundant clauses removed, $\alpha = .011$, $\lambda = 1.5$.*

***Discussion.***

The model process I investigate seems rather crude. Applying the splitting method to a real CNF-formula will in general lead to new formulae which probably do not fulfill the assumptions of homogeneity and symmetry. Start with a 3-SAT-READ-3 formula, e.g., then in the first step one has $r=2$ and $s=1$. So $m$ will even decrease in this case, at least in the first step! Apart from simplification there are, however, some arguments in favour of the simplified model. First one observes that the symmetry requirement represents a sort of worst case scenario. Because if both blocks Y and Z are equal in length before CNF-isation they will produce most new clauses and add most to the appearance of other variables. So the symmetric model will probably overestimate the tendency of the running time to explode due to branching. In this sense the appearance of bounded execution times is even more surprising. Homogeneity is an oversimplification, too, of course. But as the iteration process of the algorithm proceeds literals may well be expected to be more and more distributed according to that requirement.

Also the treatment of redundancy via two phenomenological parameters is a simplistic approach. In a more realistic approximation the two parameters $\alpha$ and $\lambda$ would depend on $n$ and $x$. The outcome of such a refined treatment needs further investigations.

Using recursion relations for averaged variables to mimic the splitting process instead of following the fate of individual formulae appears to be some sort of statistical approach. Therefore the question arises whether there is some similarity to random SAT. As pointed out in the introduction the striking difference to the phase transition found there is the fact, that here the transition curve (surface) does not separate phases of satisfiability from non-satisfiability, but areas of different efficiency of the algorithm, or stated more catchy, easy from hard problems. However, this transition has not yet been certified as universal. It depends on the solving procedure used. It would be interesting to see whether different algorithms with probably different modeling simplifications will yield similar results.

Also, I cannot exclude with absolute certainty that the jump in efficiency is an artefact of the modeling procedure. Therefore, the simulation with the aid of recursion relations should be checked against numerical experiments. By now, at least some light is shed on the question under what circumstances straightforward branching algorithms can be expected to work efficiently. Our results suggest that there might be a clear separation of easy and hard (in this sense) problems depending on the values of *m*, *n* and *k*, irrespective of the solving procedure.